**Atomic scale localization of Kohn-Sham wavefunction at SiO$_2$/4H-SiC interface under electric field, deviating from envelope function by effective mass approximation**

**Authors**
Hironori Yoshioka (吉岡裕典)[1,a], Jun-Ichi Iwata (岩田潤一)[2,3], and Yu-ichiro Matsushita (松下雄一郎)[2,3,4]

**Affiliation**
[1]Advanced Power Electronics Research Center, National Institute of Advanced Industrial Science and Technology (AIST), Tsukuba 305-8568, Japan
[2]Laboratory for Materials and Structures, Institute of Innovative Research, Tokyo Institute of Technology, Tokyo 152-8550, Japan
[3]Quemix Inc., Tokyo 103-0027, Japan
[4]National Institutes for Quantum Science and Technology, Takasaki 370-1292, Japan

Corresponding author: Hironori Yoshioka
[a]Electronic mail: hironori-yoshioka@aist.go.jp

**Abstract**

To clarify the cause of the low channel conductivity at the SiO$_2$/4H-SiC interface, the wavefunction at the SiC conduction band minimum was calculated using density functional theory under an applied electric field. We found that the wavefunction for a 4H-SiC (0001) slab tends to be localized at the cubic site closest to the interface. Importantly, because the conduction electrons are distributed closer to the interface (< 5 Å) than expected from the effective mass approximation (EMA), they are more frequently scattered by interface defects. This is expected to be the reason why the channel conductivity for the (0001) face is particularly low compared with that for other faces, such as (11$\bar{2}$0). The breakdown of the EMA for the (0001) interface is related to the long structural periodicity along the [0001] direction in 4H-SiC crystals.

# I. INTRODUCTION

Metal-oxide-semiconductor field-effect transistors (MOSFETs) with 4H silicon carbide (4H-SiC) have been commercialized for power inverters and other applications.[1] Low channel conductivity at the interface between the gate insulator and 4H-SiC, which is due to a high interface-state density and a low channel mobility, remains a problem. Although the bulk mobility in SiC (800-1,000 $cm^2V^{-1}s^{-1}$)[1-3] is comparable to that in Si (1,300-1,400 $cm^2V^{-1}s^{-1}$)[4,5], the channel mobility in SiC MOSFETs (30-130 $cm^2V^{-1}s^{-1}$)[6-12] is much lower than that in Si MOSFETs (700-800 $cm^2V^{-1}s^{-1}$).[13,14]

Carbon-related defects have been believed to be the main cause of the low channel conductivity.[7,15-22] C atoms accumulate near the interface after oxidation of the SiC surface [18,19,22-25] and create defects that form localized states near the conduction band minimum (CBM).[20,21] Several groups have performed detailed investigations into processes to reduce the amount of interfacial C atoms.[7,16,17,26,27] Using electrically detected magnetic-resonance, Higa *et al.* showed that the channel mobility increases with decreasing a C-related signal intensity.[17] $H_2$ pre-treatment,[7,26,28,29] gate insulator deposition instead of SiC surface oxidation, and NO post-annealing are effective for reducing the amount of interfacial C atoms and improving the channel conductivity. However, the channel conductivity in SiC devices is still low even after these processes, indicating that there are causes besides C-related defects.

The interface-state density increases exponentially as the CBM is approached, and is comparable to the two-dimensional density of states at the SiC CBM.[9,30-32] In other words, the interface states have a tail-like distribution from the CBM. Thus, it is expected that there is a close relationship between interfacial defect states and the CBM. Theoretical calculations have focused on the search for interfacial defect structures, but little attention has been paid to the CBM, except in a few cases.[33-36]

SiC crystals consist of a stack of Si-C bilayers, and the band gap varies depending on the stacking sequence. [1,33,37,38] 2H-SiC has a stacking sequence of ABAB… (hexagonal), and has the largest band gap (3.3 eV). 3C-SiC has a stacking sequence of ABCABC… (cubic), and has the smallest band gap (2.4 eV). 4H-SiC has a stacking sequence of ABCBABCB…, and has a band gap of 3.3 eV. The valence band maximum varies little with stacking sequence,[15] whereas the CBM decreases for longer cubic stacking sequences.[33] Because 4H-SiC has only short cubic stacking sequences (ABC or CBA), the band gap is comparable to that of the structure without cubic sequences (2H-SiC). Density functional calculations for various structures containing both cubic and hexagonal sequences have shown that the CBM is localized in low-energy cubic regions.[33,34] The same is true for structures with $SiO_2$/SiC interfaces.[35]

When investigating interfaces by density functional calculations, supercell structures with periodic slabs are commonly used.[18,19,23,35,39-42] In this approach, an electric field is unintentionally induced by the spontaneous polarization of the slab.[39,40] Thus, control of the electric field is necessary to appropriately investigate the electronic states localized near the interface under the inversion condition for n-channel MOSFETs.

The channel conductivity for the 4H-SiC (0001) face is particularly low compared with that for other faces, such as (11$\bar{2}$0).[8-12,17,43] Although explanatory factors such as a high interface-state density and the actual surface being tilted 4° from (0001) have been proposed, the reasons for the low channel conductivity have remained completely unknown for two decades. In the present study, we performed density functional calculations to investigate electronic structures localized near interfaces for different $SiO_2$/4H-SiC interface models under appropriate electric fields.

# II. CALCULATION METHODS AND STRUCTURES

Calculations were performed using a real-space density functional theory code, RSDFT.[44]

Norm-conserving pseudopotentials by the method of Troullier and Martins[45] were obtained from the Quantum ESPRESSO library[46] and used for nuclei and core electrons. The generalized gradient approximation by Perdew, Burke, and Ernzerhof was used for the exchange-correlation energy.[47] A hexagonal supercell with $a = b = 10.6$ Å and $c = 70.3$ Å was used, where the grid spacing was 0.19 Å, corresponding to a cut-off energy of 74 Ry. The $k$ points in the Brillouin zone were 2×2×2.

A $2\sqrt{3} \times 2\sqrt{3}$ R30° (0001) SiC slab was embedded in the supercell. The C dangling bonds at the $(000\bar{1})$ surface were terminated with H atoms. To form an amorphous $SiO_2$ layer, a structure with Si and O atoms scattered on the outermost Si layer of the (0001) SiC surface was prepared. It was then heated to 1,500 K and cooled to 20-10 K by the Car-Parrinello molecular dynamics simulations implemented in RSDFT.[48] The dangling bonds of the outermost atoms of the $SiO_2$ layer were also terminated with H atoms. To clearly demonstrate the effect of the stacking sequence on the CBM, we eliminated defect structures in the $SiO_2$ and at the interface that formed localized levels around the SiC band gap. For example, one 3-fold coordinated Si atom was passivated by adding one H atom. Consequently, a defect-free interface was formed between amorphous $SiO_2$ and (0001) SiC. After that, the structure was optimized until the force on the atoms was less than 0.25 eV/Å. The thicknesses of the SiC, $SiO_2$, and vacuum layers for the obtained structure were approximately 30, 13, and 25 Å, respectively. We prepared four SiC slabs with different stacking sequences as shown at the top of Fig. 1(a, b, d, and e) and Fig. 2. $4H_H$ and $4H_C$ are 4H-SiC slabs terminated with hexagonal ({A}BA) and cubic ({A}BC) sequences, respectively. A defect-free structure with a 4H-SiC $(11\bar{2}0)$ slab was also formed by the same procedure. It is noted that the CBM depends not only on the electric field but also on the slab thickness due to quantum confinement effects, especially in low electric fields. Thus, we compared slabs with the same thickness of approximately 30 Å to clarify the effect of the electric field.

The electric field was controlled by changing the charge of the H nuclei at both ends of the $SiO_2$/SiC slab.[41,42] For example, to set the electric field to 1.0 MV/cm for the $4H_H$ structure, seven H pseudopotentials at the $SiO_2$ (SiC) side were replaced by pseudopotentials with a fractional nuclear charge of $+1.0184e$ ($+0.9816e$), without changing the total nuclear charge or the electron number. Figure 3 shows the local potential ($V_{loc}$) averaged over a 2.5-Å-thick (one Si-C bilayer thick) slab for different charges. The slope of the local potential in the SiC slab is taken as the value of the electric field. We confirmed that the electric field in the SiC slab can be linearly controlled by the charge. We applied electric fields corresponding to those in actual MOSFETs.[49] A positive electric field corresponds to applying a positive voltage to the gate of a MOSFET.

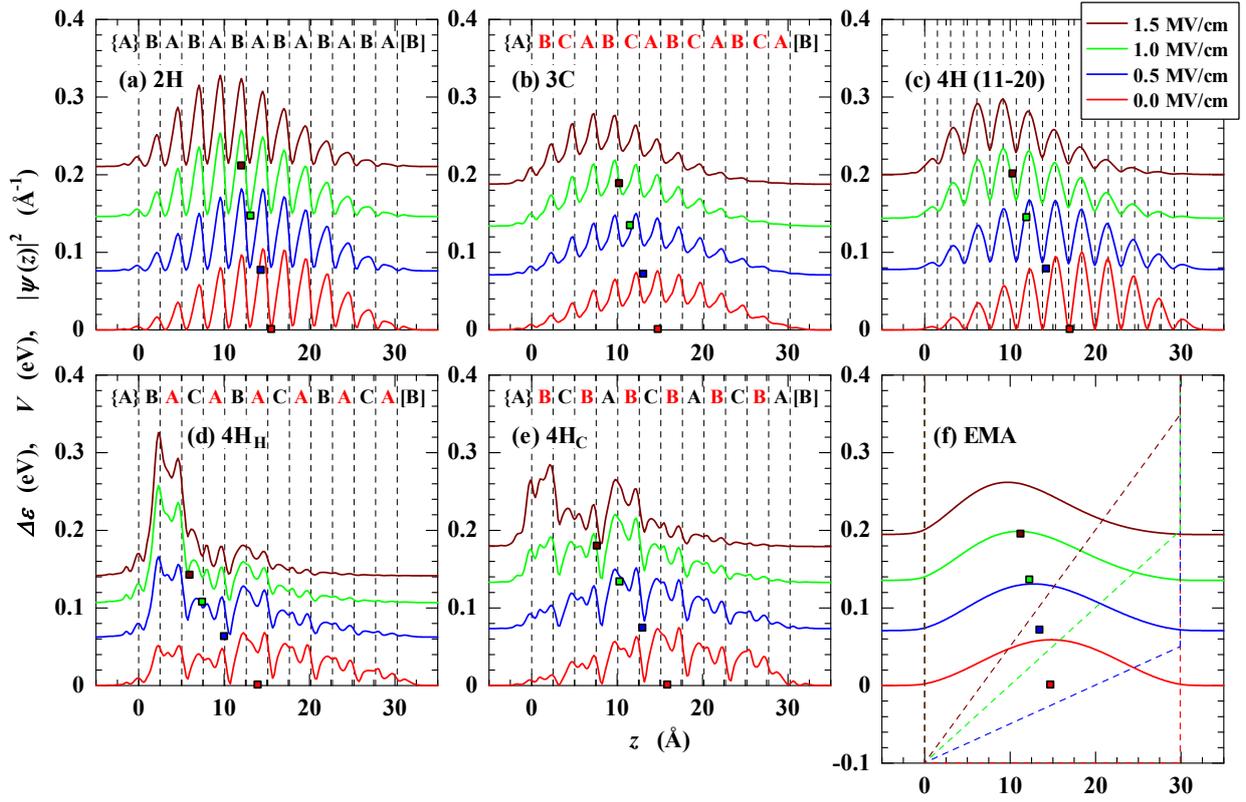

FIG. 1. Change in CBM for various SO$_2$/SiC models due to electric field ($F$= 0.0-1.5 MV/cm). (a, b, d, and e) (0001) SiC slab with different stacking sequences indicated at the top of each figure, where red letters indicate the cubic sites in the SiC crystal. (c) (11$\bar{2}$0) 4H-SiC slab. The vertical dashed lines indicate the position of the Si layers. The Kohn-Sham wavefunction ($|\psi(z)|^2$) for the CBM, integrated over the slab plane, is plotted with solid lines with an upward shift by its energy ($\Delta\varepsilon$). (f) The EMA results with $m^* = 0.33m_0$[38], where trapezoidal input potentials ($V$) are also shown by dashed lines. The energy changes due to the electric field ($\Delta\varepsilon$) versus the expectation value of the position ($\bar{z}$) is plotted with square markers.

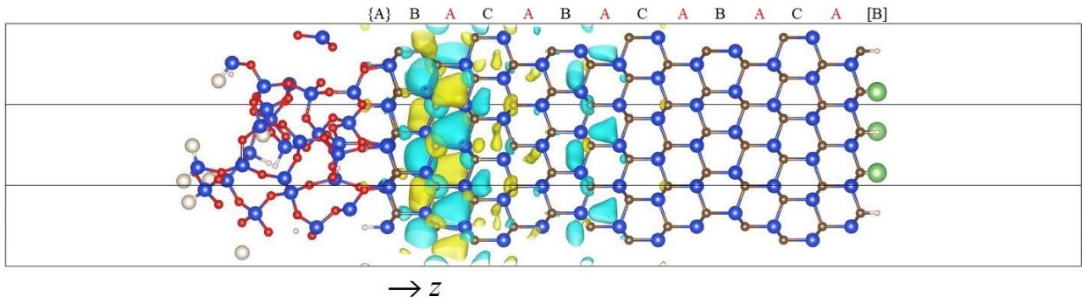

FIG. 2. Isosurface at 38% of the peak value of typical Kohn-Sham wavefunction localized at outermost cubic site (4H$_H$, $F$ = 1.0 MV/cm, CBM, $\Gamma$ point). The stacking sequence above and $z$ axis below are consistent with those in Fig. 1(d). The blue, brown, red, and small white balls depict Si, C, O, and H atoms, respectively. The large green and white balls depict H atoms with a modified pseudopotential to control the electric field.

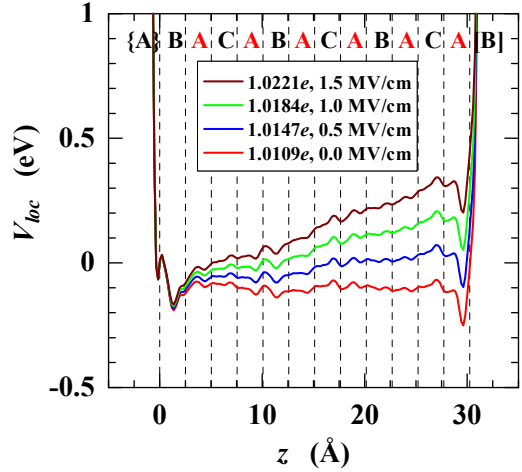

FIG. 3. Local potential ($V_{loc}$) for $4H_H$ averaged over a 2.5-Å-thick (one Si-C bilayer thick) slab. The nuclear charge for the seven H pseudopotentials on the $SiO_2$ (SiC) side changes from +1.0109$e$ (+0.9891$e$) to +1.0221$e$ (+0.9779$e$), resulting in electric fields from 0.0 to 1.5 MV/cm, as shown in the legend. $V_{loc}$ at $z$ = 0 Å is set to 0 eV.

## III. RESULTS AND DISCUSSION

Figure 1 shows the change in the CBM due to the electric field ($F$) for various $SiO_2$/SiC models. The SiC stacking sequence is shown at the top of each figure, where red letters indicate cubic sites in SiC crystals. The vertical dashed lines indicate the position of the Si layers. Let the outermost Si layer be the interface ($z$ = 0). The square markers show the energy change ($\Delta\varepsilon$) due to the electric field versus the expectation value ($\bar{z}$) of the position. The energy change is given by

$$\Delta\varepsilon = \varepsilon(F) - \varepsilon(F=0) - \{V_{loc}(F, z=0) - V_{loc}(F=0, z=0)\}, \quad (1)$$

where $\varepsilon$ is the energy level of the CBM and the term in braces {} is added to cancel the change in the local potential ($V_{loc}$) at $z$ = 0. The Kohn-Sham wavefunction for the CBM is integrated over the slab plane as

$$|\psi(z)|^2 = \int |\psi(x,y,z)|^2 \, dxdy \quad (2)$$

and plotted in Fig. 1 with an upward shift by its energy ($\Delta\varepsilon$). $\bar{z}$ is defined as

$$\bar{z} = \int |\psi(z)|^2 z \, dz. \quad (3)$$

As the electric field increases, the wavefunction approaches the interface and increases in energy. This increase in energy is mainly attributed to the increase in the local potential in SiC slabs. We found that the wavefunction at the CBM for the 4H-SiC (0001) slabs tends to localize to the cubic site closest to the interface by applying an electric field. Figure 2 shows the isosurface of the wavefunction for $4H_H$, CBM, $\Gamma$ point, and 1.0 MV/cm, which is localized at the outermost cubic site.

Figure 1(f) shows the effective mass approximation (EMA) results for solving the one-dimensional effective mass equation with $m^* = 0.33m_0$[38] using Snider's code[50], where the trapezoidal input potential ($V$) is also shown. The potential gap at $z$ = 0 Å (30 Å) is set to 2.7 eV (3.6 eV), supposing a $SiO_2$/4H-SiC (4H-SiC/vacuum) interface.[15,51] It is noted that, for the case of no electric field ($F$ = 0), the CBM for a 30-Å-thick slab is 0.10 eV higher than that for an infinitely thick slab (bulk) due to quantum confinement effects, as shown in Fig. 1(f). It is obvious that the EMA is applicable to 2H, 3C, and 4H($11\bar{2}0$) but not to $4H_H$ and $4H_C$. Compared with the EMA results, the wavefunction for $4H_H$ and $4H_C$ is highly modulated and distributed very much closer to the interface (< 5 Å) than expected from the EMA, and $\Delta\varepsilon$ is smaller. The breakdown of the EMA is an important finding, because the EMA is commonly used to investigate carrier transport.[49,52-55] The EMA is applicable only when the perturbation potential is approximately constant over a primitive unit cell.[56] Since the 4H-SiC (0001) slab has a very long periodicity along [0001] (10.0 Å), the breakdown of EMA is reasonable.[57]

The more the CBM wavefunction is localized near the interface, the more the CBM carriers are scattered by scattering factors at the interface, resulting in a decrease in relaxation time ($\tau$) and mobility ($\mu$) ($\mu = e\tau/m^*$). From another viewpoint, the more the CBM wavefunction is localized near the interface, the more the CBM states are changed with mixing with the interface defect states, forming a tail-like density of states. Carriers in such tail states are expected to have a large effective mass and small mobility, and have actually been detected because they exhibit hopping conduction at cryogenic temperatures.[32]

The localization is more conspicuous for the 4H-SiC (0001) face [Fig. 1(d, e)] than that for the ($11\bar{2}0$) face [Fig. 1(c)], which is expected to be the reason why the channel conductivity for the (0001) face is particularly low compared with that for the ($11\bar{2}0$) face.[8-12,17,43] Hirai *et al.* clearly showed that the difference in Hall mobility between the faces becomes significant as the electric field increases,[12] which is consistent with our results on localization of the wavefunction. It is noted that, although so called (0001) for actual devices is slightly tilted (typically 4°), the interface forms a step-terrace structure containing large (0001) terraces.

## IV. CONCLUSION

We found that the wavefunction at the CBM for 4H-SiC (0001) slabs tends to be localized at the cubic site closest to the interface when an electric field is applied. Because the conduction electrons are highly concentrated very close to the interface (< 5 Å), they are more frequently scattered by interface defects, which is expected to be the reason why the channel conductivity for the (0001) face is particularly low compared with that for other faces, such as ($11\bar{2}0$). The breakdown of the EMA for the 4H-SiC (0001) interface is related to the long crystal periodicity perpendicular to the slab.

**ACKNOWLEDGMENTS**

The computation in this work has been done using the facilities of the Supercomputer Center, the Institute for Solid State Physics, the University of Tokyo. This work was supported by JSPS KAKENHI Grant Number 21H04553, 20H00340, and 20K05352. This work was partly supported by MEXT as "Program for Promoting Researches on the Supercomputer Fugaku" (JPMXP1020200205). The computation in this work has been done partly using supercomputer Fugaku provided by the RIKEN Center for Computational Science. VESTA,[58] Sma4 and Google Translate are used to prepare the manuscript.

The data that support the findings of this study are available from the corresponding author upon reasonable request.